\documentclass[12pt]{article}
\usepackage{amsmath,amsfonts,amssymb,bm}
\usepackage{graphicx}
\usepackage{natbib}
\usepackage{url} 
\usepackage{xcolor}
\usepackage{diagbox}

\usepackage{caption}
\usepackage{subcaption}

\newcommand{\blind}{0}

\newcommand{\mb}{\mathbf}
\newcommand{\bs}{\boldsymbol}

\begin{document}

\def\spacingset#1{\renewcommand{\baselinestretch}%
{#1}\small\normalsize} \spacingset{1}


\if0\blind
{
  \title{\bf \vspace{-1.5cm} A Neural Network-Based Approach to Normality Testing for Dependent Data}
  \author{Minwoo Kim, \\
    \small Department of Statistics, Pusan National University, \\
    Marc G. Genton,  Rapha\"el Huser,\\
    \small Statistics program, King Abdullah University of Science and Technology, \\
    and Stefano Castruccio\thanks{
    Corresponding author, scastruc@nd.edu}\hspace{.2cm} \\
    \small Department of Applied and Computational Mathematics and Statistics, 
    University of Notre Dame } 
  \maketitle
} \fi

\if1\blind
{
  \bigskip
  \bigskip
  \bigskip
  \begin{center}
    {\LARGE\bf Title}
\end{center}
  \medskip
} \fi
\vspace{-1cm}
\begin{abstract}
There is a wide availability of methods for testing normality under the assumption of independent and identically distributed data. When data are dependent in space and/or time, however, assessing and testing the marginal behavior is considerably more challenging, as the marginal behavior is impacted by the degree of dependence. We propose a new approach to assess normality for dependent data by non-linearly incorporating existing statistics from normality tests as well as sample moments such as skewness and kurtosis through a neural network. We calibrate (deep) neural networks by simulated normal and non-normal data with a wide range of dependence structures and we determine the probability of rejecting the null hypothesis. We compare several approaches for normality tests and demonstrate the superiority of our method in terms of statistical power through an extensive simulation study. A real world application to global temperature data further demonstrates how the degree of spatio-temporal aggregation affects the marginal normality in the data. 
\end{abstract}

\noindent%
{\it Keywords:} Adaptive Cut-Off; Aggregation of Test Statistics; Neural Network; Normality Test; Spatio-Temporal Statistics

\vspace{-.5cm}
\spacingset{1.5} 
\section{Introduction}
\label{sec:intro}
One of the fundamental tasks for both model design and validation is to identify a marginal distribution for the data (or the residuals according to some trend), and to test whether it can be ascribed to a known parametric model. Arguably one, if not the the most, important case is that of the normal distribution. In this case, in addition to informal methods such as quantile-quantile plots and histograms, there is a wide variety of normality tests under the assumption of independent and identically distributed (\textit{i.i.d.}) data; see, e.g., \cite{and52}, \cite{sha65}, \cite{lil67}, and \cite{jar80}. Normality tests are based on statistics such as skewness and kurtosis, which summarize some properties of the distribution and compare them to the statistic expected from a normal distribution. The tests may not provide unanimous results if, for instance, the data resemble a normal distribution with respect to one statistic but not with respect to others; see \citet{thode2002testing}. 

When the data are not \textit{i.i.d.}, with dependence informed possibly (but not necessarily) by space and/or time, testing the marginal behavior is considerably more challenging. Indeed, while it is methodologically convenient to assume a Gaussian process, i.e., a random function with marginal Gaussian distribution, the dependence leads to excessive rejections in normality tests intended for \textit{i.i.d.}\ data. As an extreme example, one may consider a Gaussian process with perfect correlation: for every realization, every observation will have the same value, hence leading to the impossibility of assessing the marginal behavior. Therefore, standard tests intended for \textit{i.i.d.}\ data are bound to exhibit inflated Type I error rates on dependent data, even if the process is in fact Gaussian. It is hence necessary to develop tests that account for dependence, and which would adjust the decision criterion accordingly. The recent work of  \citet{hor20} proposed a modification of the Jacque-Bera normality test \citep{jar80} by estimating the spatial structure. In their review on multivariate normality tests, \citet{chen2022you} also extended the test of \citet{hor20} to the multivariate setting.

While a test adjustment may provide a partial solution, relying on  only a single test with dependent data is limiting, as the null distribution of the test statistic strongly depends on the correlation structure. For instance, the null distribution of a single test statistic such as the Shapiro-Wilk normality test \citep{sha65} will differ depending on the strength of the spatial dependence. In order to enhance the test power, a solution is to combine different tests so as to use multiple statistics at the same time.

One simple approach is the Bonferroni correction, which predicates rejection of $H_0$ if at least one of the $m$ tests is rejected at level $\alpha / m$; see, e.g., \citet{hay13}. The Bonferroni correction guarantees the appropriate Type I error rate but is overly conservative and has an optimal power only if the test statistics are independent. Another approach to combine $m$ tests is to use Fisher's method, which combines information from the p-values of all tests. If the tests are all independent, then $-2\sum_{i=1}^m \ln p_i$ follows a $\chi_{2m}^2$ distribution \citep{fisher1992statistical, kost2002combining}. A linear combinations of p-values has also been suggested in \citet{edg72}.  \citet{wink16} reviewed fifteen methods for combining p-values.

Neural networks-based approaches with descriptive statistics as inputs for \textit{i.i.d.}\ data have been introduced to test for normality and compared with standard tests \citep{wil90}. \citet{sig06} assessed univariate normality using trained neural networks with input features including sample skewness, sample kurtosis, test statistics in \citet{sha65}, the Fisher transform of the Pearson correlation coefficient, and the family of test statistics proposed by \citet{vasicek1976test}. More recently, \citet{sim20} extended previous approaches by adding summary statistics such as minimum, maximum, and sample size to the representative input set. All the past studies showed that neural network approaches can often outperform typical statistical tests by combining information in a non-linear fashion. In this work, we propose a more general neural network-based test for normality aimed at dependent data (in space, time, space/time, or simply multivariate) with a novel adaptive cut-off technique, which will be shown to outperform currently available methods for testing normality when the independence assumption is violated. 

The paper proceeds as follows. In Section \ref{sec:methods}, we present the general framework of combining multiple tests and introduce our neural network methodology. In Section \ref{sec:sim}, we conduct a simulation study for testing the assumption of normality on a spatial grid and we show the improvement against currently available methods. In Section \ref{sec:app}, we apply the proposed method to spatially distributed data from a global climate model simulation in order to test normality at different levels of spatial aggregation. In Section \ref{sec:concl}, we discuss conclusions and directions for future research.  
\vspace{-.7cm}

\section{Methodology for Normality Testing}\label{sec:methods}

Let $\bm Y = \left( Y(\mb{s}_1), \dots, Y(\mb{s}_M) \right)^\top$ be a vector of real valued random processes on a manifold. This manifold can represent a spatial domain such as a Euclidean space or a sphere for a spatial process, the positive real line for time series or a Cartesian product of the two in the case of space-time processes. Let $H_0$ be any model property that $Y(\cdot)$ may satisfy (in our case the marginal distribution being Gaussian). We aim to create a most-powerful classifier $C:  \bm Y  \mapsto \{0,1\}$ with Type I error rate $\alpha$; that is we have $P(C(\bm Y) = 1 \mid H_0 \quad \text{true}) = \alpha$ and for any other classifier $\tilde{C}$ at the same Type I error rate we have that 
$
P(C(\bm Y) = 1 \mid H_0 \quad \text{false}) \geq P(\tilde{C}(\bm Y) = 1 \mid H_0 \quad \text{false})$.

\subsection{Individual normality tests}\label{sec:indiv:tests}

For simplicity of notation, we denote with $Y_i=Y(\mb{s}_i)$, $i=1, \ldots, M$, $\bm Y=(Y_1, \ldots, Y_M)^\top$ the data for which one wants to assess normality. We focus on four tests that are used as inputs for our neural network: Shapiro--Wilk \citep{sha65}, Lilliefors \citep{lil67}, Jarque--Bera \citep{jar80}, and Anderson--Darling \citep{and52}.

The Shapiro--Wilk test relies on calculating the order statistics and comparing the observed versus expected values
$W = (\sum_{i=1}^M a_i Y_{(i)})^2/\sum_{i=1}^M (Y_i - \bar{Y})^2$,
where \(Y_{(i)}\) is the $i^{th}$ order statistic, $\bar{Y}$ is the sample mean, and $a_i$ is a weight calculated from the expected means and covariances of the order statistics under the null hypothesis of \textit{i.i.d.}\ data. Despite its popularity, the Shapiro--Wilk test relies on the availability of appropriate values of $a_i$ which have no closed form, so the values are determined through Monte Carlo simulation, and for large sample sizes $M$, it is more difficult to obtain accurate $a_i$ estimates \citep{das2016brief}. Indeed, in all the code implementation we used throughout this work, the size of $M$ is limited to a few thousand points. 

The Lilliefors test is an adaptation of the Kolmogorov--Smirnov test for Gaussian data. It measures the maximum deviation of the empirical and theoretical cumulative distribution functions (CDFs), denoted with $F_M$ and $F$, respectively:
$D_M = \sup_y |F_M(y) - F(y)|$.
Then $D_M$ is compared to the expected distribution under the null hypothesis, and a p-value is calculated.

The Anderson--Darling test statistic is also based on deviation from the theoretical CDF:
$A^2 = M \int_{-\infty}^{\infty} \frac{\{F_M(y) - F(y)\}^2}{F(y) \{1-F(y)\}}\mathrm{d}F(y)$.
Rather than measuring the maximum deviation between the empirical and theoretical CDFs, Anderson--Darling weighs deviations in the tails more heavily.  

Finally, the Jarque--Bera test calculates the test statistic
$JB = \frac{M}{6} \{S^2 +(K-3)^2/4\}$,
where $S$ and $K$ are the sample skewness and kurtosis, respectively. Informally, the Jarque–Bera test checks whether the sample’s skewness and kurtosis match those of a normal distribution. The asymptotic expected values of the empirical skewness and kurtosis are 0 and 3, and the asymptotic variance of the empirical skewness and kurtosis are $6/M$ and $24/M$. Thus, the Jarque--Bera statistic is a squared sum of two asymptotically independent standardized normal distributions, and thus distributed as a $\chi^2$ random variable.  
\vspace{-.3cm}

\subsection{Combining tests}

Let $C_1, C_2, \ldots, C_m$ be $m$ classifiers with Type I error $\alpha$. Insofar as they are distinct classifiers, they assess at least partly different properties implied by $H_0$. For example, to test $H_0$: $\bm Y({\bf s})$ is normally distributed, $C_1$ may be testing whether the skewness is zero, while $C_2$ may be testing whether the excess kurtosis is zero. Both are appropriate level-$\alpha$ tests of $H_0$ and their performance, measured by statistical power, will vary depending on how the departure of the alternative model hypothesis $H_1$ to $H_0$ affects the properties assessed by each classifier.  

Ideally, we would like to combine the $m$ classifiers into a single level-$\alpha$ classifier $C$ that is more powerful. In our case, combining the classifiers is complicated because of two main issues. First, since each individual classifier is testing different but related properties of $H_0$, the $m$ classifiers are expected to be dependent; the Bonferroni correction is overly conservative because the effective number of tests is less than $m$ due to this dependence and Fisher's method's asymptotic distribution is no longer valid. In the field of statistical genetics, \citet{gre15} accounted for the dependence of various genetic tests of association for case-control studies by repeatedly permuting cases and controls in order to calculate the null distribution of either the Fisher statistic or minimum p-value statistic, which naturally adjusts for the dependence. The method relies on creating a representative sample of data under the null hypothesis through permutations. In our setting, we only have a single realization of the process $\bm Y({\bf s})$, so instead of a permutation, we will create a representative sample of data under $H_0$ through simulation.

\subsection{Combining tests through neural networks}

If $T_1, T_2, \ldots, T_m$ are test statistics for classifiers $C_1, C_2, \ldots, C_m$, the simplest approach to combine them is through a classifier  comprising of a linear combination and a logit transformation: $
\text{logit}\{P(C(\bm Y) = 1)\} = \gamma_0 + \gamma_1 T_1 + \cdots + \gamma_m T_m$.
While this approach allows to combine information across tests, its functional form limits its flexibility. In this work, we propose a more flexible approach which relies on a (deep) neural network, i.e., we filter the test statistics through a combination of multiple non-linear functions \citep{goo16}. More specifically, we consider the following: 
\begin{align}
\label{eq:NN}
    F(\bm Y)=P(C(\bm Y) = 1) = S\{W_L \sigma_L (W_{L-1} \cdots \sigma_2(W_2 \sigma_1 (W_1 \bm T)))\},    
\end{align}
which is a composition of:
\begin{enumerate}
\item The $m$-dimensional vector of all the test statistics considered $\bm T=(T_1,\ldots, T_m)^\top$. If no classifiers are available, one may also consider $\bm T$ to be the identity function so that the vector of the observed data $\bm Y$ itself is the desired input. For simplicity of notation in the next points, we set $n_0=m$.

\item $L$ matrices representing linear transformations $W_i: \mathbb{R}^{n_{i-1}} \mapsto \mathbb{R}^{n_i}$. The parameter $n_i$ is the \textit{width} of layer $i$, while $L$ is the \textit{depth} of the neural network.

\item $L$ fixed non-linear transformations $\sigma_i$ that are applied component-wise. In this paper, we use the common restricted linear unit (ReLU, \citet{goo16}) activation function defined by $\sigma(z) = \max(0, z)$. 

\item A sigmoid function $S(z) = (1+e^{-z})^{-1}$, which guarantees an output in $[0,1]$ that we can interpret as $P(C(\bm Y) = 1)$.
\end{enumerate}

Inference (i.e., learning) can be performed by simulating the representative samples $\bm Y_1^{H_0}, \ldots, \bm Y_{N_0}^{H_0} \in \mathbb{R}^M$ satisfying $H_0$ and $\bm Y_1^{H_1}, \ldots, \bm Y_{N_1}^{H_1}$ satisfying $H_1$. The matrix entries of $W_i$ are then learned by optimizing the binary cross-entropy (or log loss), which penalizes overly-confident incorrect predictions: if we denote by $p^{H_0}_i=P\left(C\left(\bm Y_i^{H_0}\right)=1\right)$ and $p^{H_1}_i=P\left(C\left(\bm Y_i^{H_1}\right)=1\right)$, then:\\
\begin{equation}\label{eq:crossent}
\text{logloss} = \sum_{i=1}^{N_0} \log \left(1-p^{H_0}_i\right) + \sum_{i=1}^{N_1} \log p^{H_1}_i.
\end{equation}

In this work, we use the stochastic gradient descent-based optimization algorithm Adam \citep{kin15}. Since the neural network outputs a probability, instead of setting an arbitrary cut-off of 0.5, we set it such that the method has a pre-specified Type I error rate $\alpha$. Formally, this cut-off $q_{\alpha}$ is defined using \eqref{eq:NN} as:
\begin{equation}\label{eq:alpha}
q_{\alpha} = \inf_{q\in [0,1]}  \left[\frac{1}{N_0} \sum_{i=1}^{N_0} \mathbb{I}\{F(\bm Y_i^{H_0}) > q\} \leq \alpha\right].
\end{equation}

\subsection{Adaptive cut-off}

In this section we assume for simplicity that the Gaussian training data are spatially dependent and generated from a Mat\'ern covariance model \citep{ste99} with varying degrees of spatial dependence. The proposed adaptive cut-off approach can however be easily generalized to other spatial, temporal and spatio-temporal models. For any two observations $Y(\mb{s}_i), Y(\mb{s}_j)$ at two generic locations $\bs{s}_i, \bs{s}_j\in \Bbb{R}^2$, the covariance in the Mat\'ern model is:
\begin{equation}\label{eq:matern}
\text{cov}\{Y(\mb{s}_i), Y(\mb{s}_j)\}= \frac{\sigma^2}{ 2^{\nu-1}\Gamma(\nu)}\left(\frac{\|\bs{s}_i-\bs{s}_j\|}{\beta}\right)^{\nu} {\cal K}_{\nu}\left(\frac{\|\bs{s}_i-\bs{s}_j\|}{\beta}\right),
\end{equation}
where ${\cal K}_\nu$ is the modified Bessel function of the second kind of order $\nu > 0$, and $\|\bs{s}_i-\bs{s}_j\|$ is the Euclidean distance. The parameter $\sigma^2$ specifies the marginal variance and $\beta > 0$ controls the range of the spatial dependence: when we consider a distance $\sqrt{8\nu}/\beta$, the spatial correlation is near 0.1 for all $\nu$ \citep{ste99}. Finally, $\nu$ specifies the regularity/smoothness of the process, i.e., the degree of mean square differentiability. 

Since we simulate the training data by varying the spatial range $\beta$, a single cut-off value independent of this parameter would inevitably result in incorrect Type I error rates. In this work, we propose a more flexible cut-off $q_\alpha$ in \eqref{eq:alpha} as a function of $\beta$. Specifically, let $n_{\beta;\text{train}}$ be the number of range parameters for the training set such that  $\beta_1, \dots, \beta_{n_{\beta;\text{train}}}$ are the parameters used to generate $\bm Y_1^{H_0}, \dots, \bm Y_{N_0}^{H_0}$. For each $\beta_g$ and its corresponding observations, a cut-off value is elicited as in \eqref{eq:alpha} denoted by $q_{\alpha}(\beta_g)$ for $g=1, \dots, n_{\beta;\text{train}}$. We employ non-parametric kernel regression to estimate the cut-off function based on $n_{\beta;\text{train}}$ pairs $(\beta_1, q_{\alpha}(\beta_1))^\top, \dots, (\beta_{n_{\beta;\text{train}}}, q_{\alpha}(\beta_{n_{\beta;\text{train}}}))^\top$. We use a Gaussian kernel and assume that the estimated cut-off at a new testing value $\beta$ is:
\begin{equation}\label{eq:kernel}
    \hat{q}_{\alpha}(\beta) = \frac{\sum_{g=1}^{n_{\beta;\text{train}}} K_h(\beta - \beta_g)q_{\alpha}(\beta_g)}{\sum_{g=1}^{n_{\beta;\text{train}}} K_h(\beta - \beta_g)},
\end{equation}
where $K_h(\beta - \beta_g) = h^{-1} K\left( h^{-1}(\beta - \beta_g) \right)$, $K(z) = \exp \left( - z^2 / 2 \right) / \sqrt{2\pi}$ for any $z \in \mathbb{R}$, and $h$ is a selected bandwidth. We implement this kernel regression using the R package \texttt{np} \citep{li2003nonparametric, li2013optimal}. 

\subsection{An existing test for dependent normal data}

\citet{hor20} introduced a test to determine whether some dependent data on a regular grid can be regarded as a realization of a Gaussian process. We show here the main idea behind their approach, and we refer to their manuscript for a comprehensive derivation of the test statistic and relevant estimators. Their method involves modeling a process that accounts for the spatial correlation and computing two statistics related to sample skewness and kurtosis. The test can be performed since \citet{hor20} demonstrated that the sum of squares of the two statistics asymptotically follows a chi-square distribution with two degrees of freedom. Specifically, the data $\left\{Y(\mb{s}_1), \dots, Y(\mb{s}_M)\right\}$, where $\{\mb{s}_1, \dots, \mb{s}_M \} \in \mathbb{Z}^d$ are locations in a $d$-dimensional spatial domain, are assumed to follow the moving average model
$Y(\mb{s}) = \mu + \sum_{\mb{s}' \in \mathbb{Z}^d} a(\mb{s}') \epsilon(\mb{s} - \mb{s}')$,  $\mb{s} \in \mathbb{Z}^d$,
where $\mu$ is the process mean and $\epsilon(\mb{s}), \mb{s} \in \mathbb{Z}^d$ are independent, standard normal innovations. We denote sample skewness and kurtosis with the standardized data by $\mathcal{S}_M$ and $\mathcal{K}_M$ respectively, and by $\phi_\mathcal{S}^2$ and $\phi_\mathcal{K}^2$ their asymptotic variances (which depend on $\mu$ and $a(\mb{s}')$). The test statistic is defined as $\mathcal{S}_M^2/\hat{\phi}_\mathcal{S}^2 +\mathcal{K}_M^2/\hat{\phi}_\mathcal{K}^2$,
where $\hat{\phi}_\mathcal{S}^2$ and  $\hat{\phi}_\mathcal{K}^2$ are kernel estimators whose detailed explanation and comprehensive derivations are given in their paper. In Section \ref{sec:sim} of this work, we use this test as a benchmark to compare the performance of our proposed method.
\vspace{-1.4cm}

\section{Simulation Study}\label{sec:sim}
\vspace{-.5cm}

\subsection{Simulation design}
We simulate a zero mean, isotropic Gaussian random field with Mat\'ern covariance function in \eqref{eq:matern} on a two dimensional unit square regular grid of size $60 \times 60$. We assume $\nu \in \{0.5, 1.0 \}$ where the former value  simplifies the covariance function to $\sigma^2\exp(-\|\bs{s}_i-\bs{s}_j\|/\beta)$. We present results for $\nu=0.5$ in this section, while the results for $\nu=1.0$ are deferred to the supplement Section A. We choose $n_{\beta;\text{train}}=30$ equally spaced values of $\beta$ between 0 and $\beta_{\max}=0.234$ (including both endpoints) in the training set, spanning from zero to strong dependence on a unit square. The range parameter bound $\beta_{\max}$ is chosen so that the \textit{effective range}, i.e., the distance at which the correlation between two locations reaches 0.05, is 0.7. This bound is valid only for the unit square, so it requires a rescaling in the application, and also depends on $\nu$. In the test set we choose $n_{\beta;\text{test}}=50$ equally spaced values of $\beta$ from 0 to $\beta_{\max}$, to demonstrate that the neural network is capable of interpolating between different choices of range parameters. The sets of $\beta$s in the training set and testing set are denoted by $\mathcal{B}_{\text{train}}$ and $\mathcal{B}_{\text{test}}$, respectively, such that $\lvert \mathcal{B}_{\text{train}} \rvert = n_{\beta;\text{train}}$ and $\lvert \mathcal{B}_{\text{test}} \rvert = n_{\beta;\text{test}}$. Non-normal distributions in the training and testing set were created by applying a signed power transformation to the baseline Mat\'ern Gaussian random field. Specifically, for an exponent parameter $p$, a value $z$ was transformed to $f(z;p) = \lvert z \rvert^p \text{sign}(z)$, for values of $p$ in the set $\mathcal{P}_{\text{train}} = \{1.2, 1.4, 1.6, 1.8\}$ in the training set, and in the set $\mathcal{P}_{\text{test}} = \{1.1, 1.2, \dots, 2.0\}$ in the testing set, to demonstrate the neural network's ability to interpolate and (modestly) extrapolate. We denote by $\lvert \mathcal{P}_{\text{train}} \rvert = n_{p;\text{train}}$ and $\lvert \mathcal{P}_{\text{test}} \rvert = n_{p;\text{test}}$, and we generate $n_{\text{sample}} = 200$ sample points for each combination of ($\beta, p)$ in the case of non-normal data. Therefore, the training set contains $n_{\beta;\text{train}} \times n_{p;\text{train}} \times n_{\text{sample}} = 24,000$ (non-normal) data points, while the testing set contains $n_{\beta;\text{test}} \times n_{p;\text{test}} \times n_{\text{sample}} = 100,000$ (non-normal) data points. For the null hypothesis, i.e., normal data with $p=1$, we generate an equivalent number of samples, i.e, the training set contains $24,000$ points, while the testing set contains $100,000$ points using the same sets $\mathcal{B}_{\text{train}}$ and $\mathcal{B}_{\text{test}}$, respectively.

Type I errors for individual normality tests introduced in Section \ref{sec:indiv:tests} are presented in Section \ref{sim:classic}. Results in terms of Type I error and power for our neural network, the linear classifiers and \citet{hor20}'s method are shown in Section \ref{sim:result}.
\vspace{-.3cm}

\subsection{Classical tests} \label{sim:classic}
\begin{figure}[t!]
\centering
\includegraphics[scale=0.34]{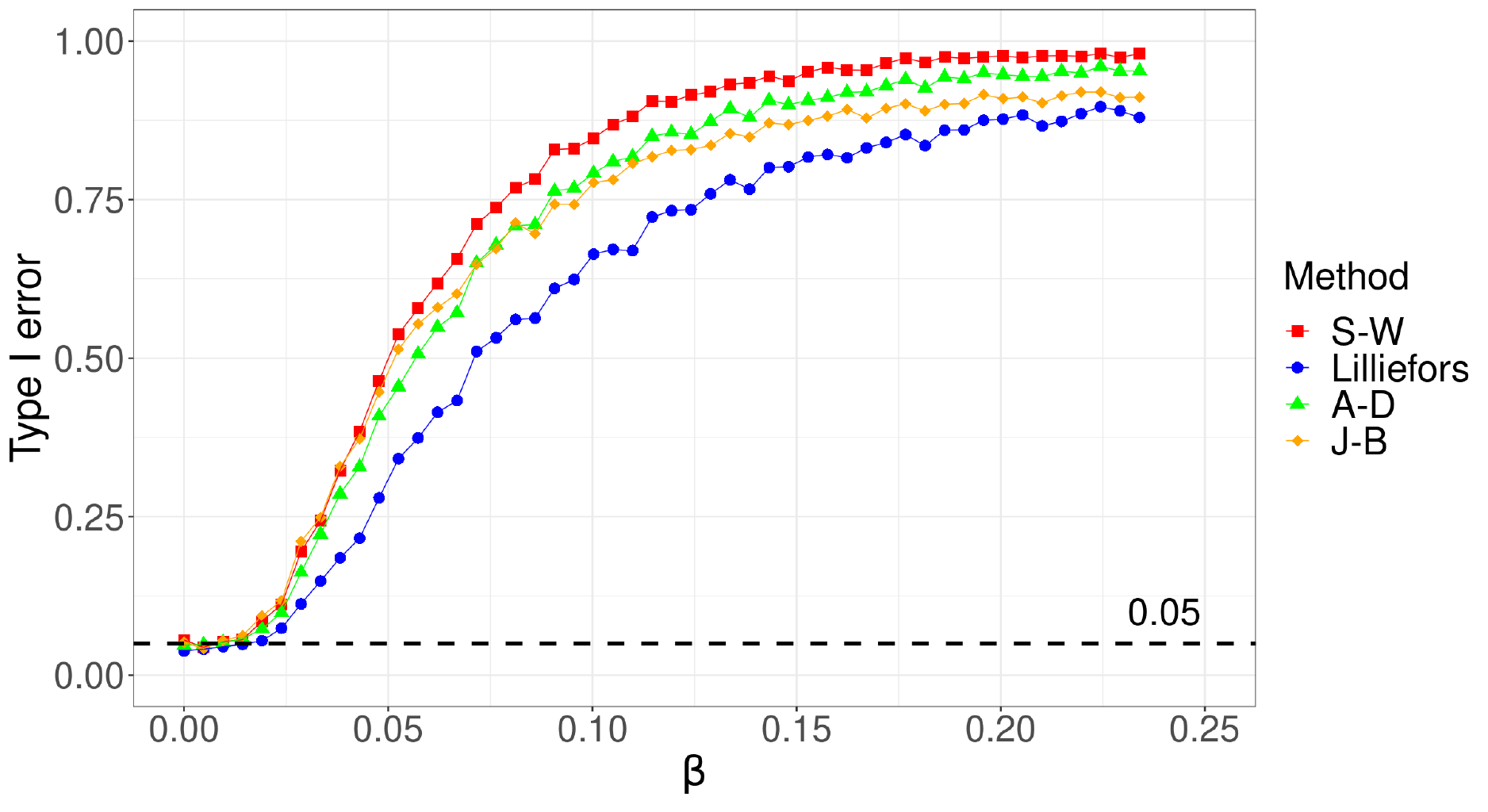}
\caption{ \small $y$-axis: Type I errors for Shapiro--Wilk test (red), Lilliefors test (blue), Anderson--Darling test (green), and Jarque--Bera test (orange). $x$-axis: The dependence parameter $\beta$ of a Mat\'ern covariance function as shown in \eqref{eq:matern} when the other two parameters are fixed, $\sigma^2 = 1$ and $\nu = 0.5$. The black dashed horizontal line in the figure represents 5\% of Type I error.}
\label{type1err:classic}
\end{figure}

The Type I errors for the classical normality tests increase as the range of dependence increases in the simulation data, as is apparent in Figure \ref{type1err:classic}. These tests are therefore not appropriate given their assumption of independence. Given their uncalibrated Type I error, we do not calculate the power of these tests and do not compare them with the other methods shown in the following sections. 

\subsection{Tests for dependent data} \label{sim:result}

\begin{figure}[t!]
\centering
\includegraphics[scale=0.34]{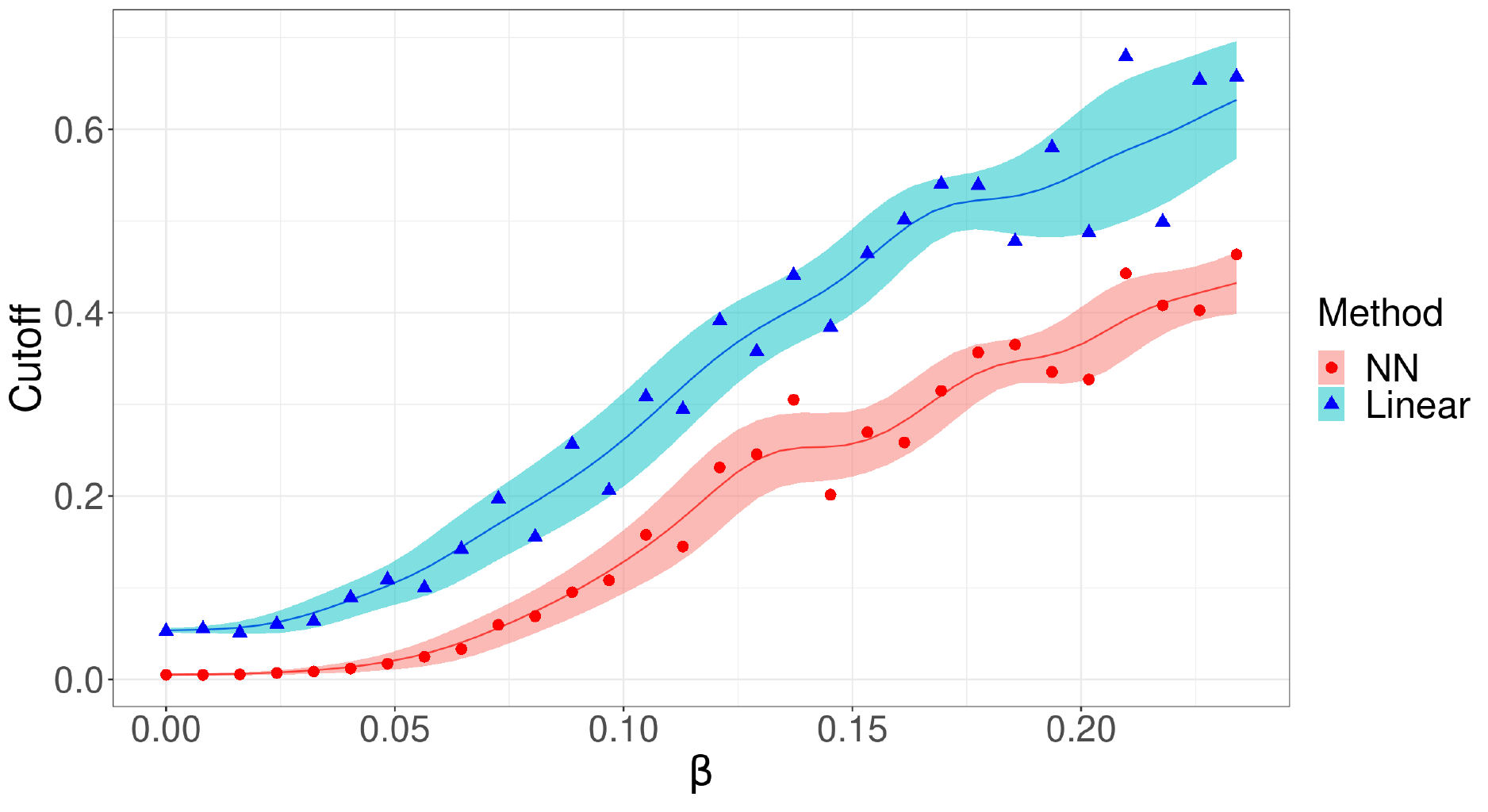}
\caption{\small Simulation study: Non-parametric Gaussian kernel regressions as defined in \eqref{eq:kernel} with a bandwidth $h=0.3$ for neural network (red) and linear (blue) classifiers. On the $x$-axis are the range parameter $\beta$ of the Mat\'ern covariance \eqref{eq:matern}, while on the $y$-axis the predicted cut-off and corresponding pointwise 95\% confidence interval are represented by solid lines and bands, respectively. The other two parameters are fixed at $\sigma^2 = 1$ and $\nu = 0.5$.}
\label{cutoff:nu05}
\end{figure}

We use $m=6$ inputs: the four test statistics of the normality tests in Section \ref{sec:indiv:tests} along with the sample skewness and kurtosis. We rely on a neural networks with $L=2$ hidden layers and with $n_1=256$ and $n_2=128$ nodes. To at least partly mitigate overfitting we use dropout \citep{sriv14} during training, which randomly removes a fraction of nodes during each training step and acts as a form of regularization. In each of the $L$ layers, 30\% of nodes are randomly removed during each training step. We provide a sensitivity study in Section \ref{sim:sensitivity} to demonstrate the robustness of the results with respect to other choices of network depth, width and drop-out rate. Inference is performed by minimizing the binary cross-entropy logarithmic loss \eqref{eq:crossent}, which is equivalent to maximizing the log-likelihood. For each $\beta \in \mathcal{B}_{\text{train}}$, we set a cut-off at the observed $1-\alpha=95$th percentile in \eqref{eq:alpha} using the associated Gaussian data in training set such that we collect $(\beta_1, q_{\alpha}(\beta_1))^\top, \dots, (\beta_{n_{\beta;\text{train}}}, q_{\alpha}(\beta_{n_{\beta;\text{train}}}))^\top$ and obtain cut-off functions for neural network and linear classifiers from non-parametric kernel regression as shown in Figure \ref{cutoff:nu05}.

\subsubsection{Type I error comparison}

\begin{figure}[b!]
    \centering
    \begin{subfigure}[t]{0.49\textwidth}
        \centering
        \includegraphics[height=5cm, width=7.2cm]{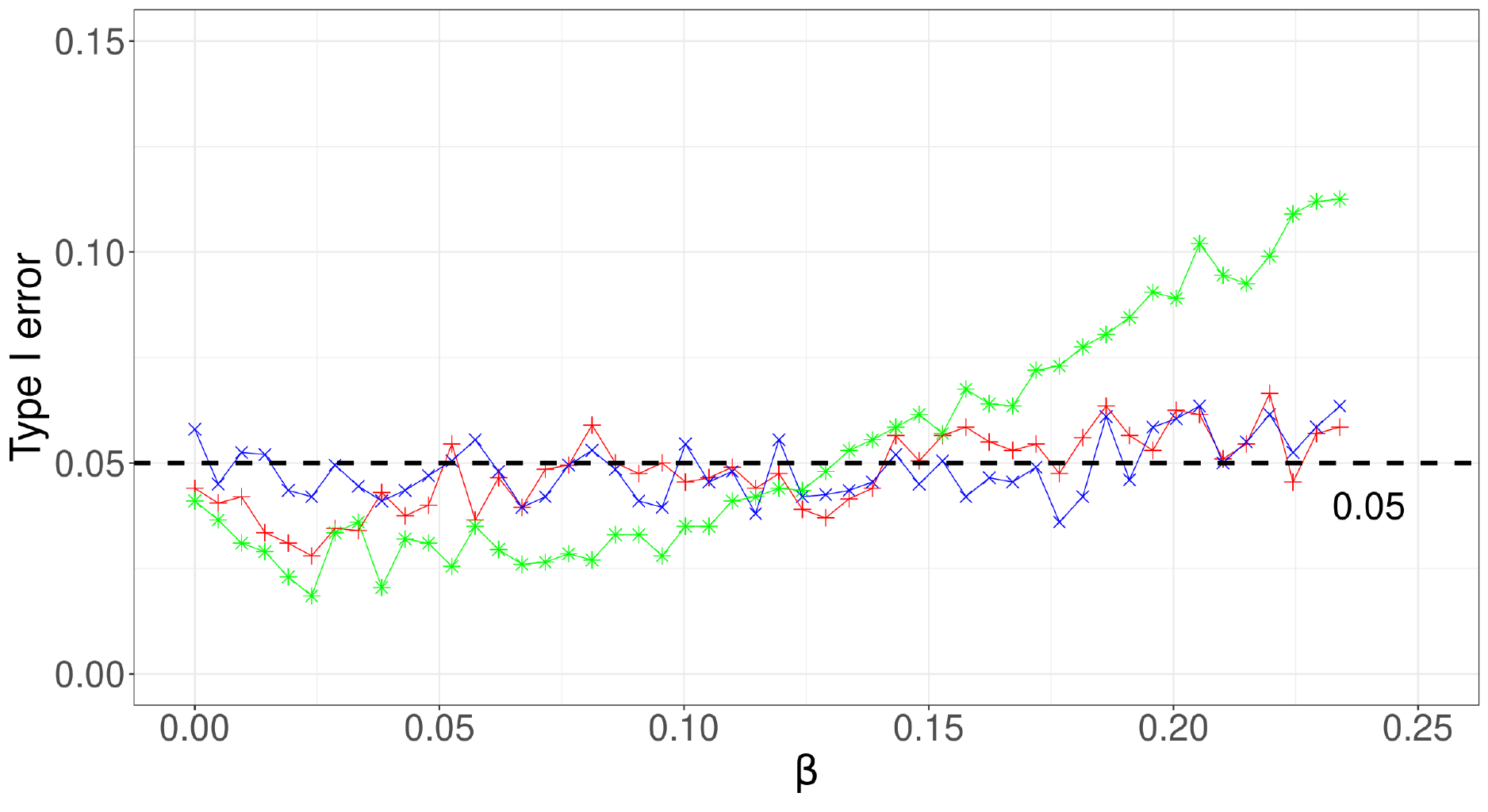}
        \caption{}
    \end{subfigure}%
    ~
    \begin{subfigure}[t]{0.49\textwidth}
        \centering
        \includegraphics[height=5cm, width=8.5cm]{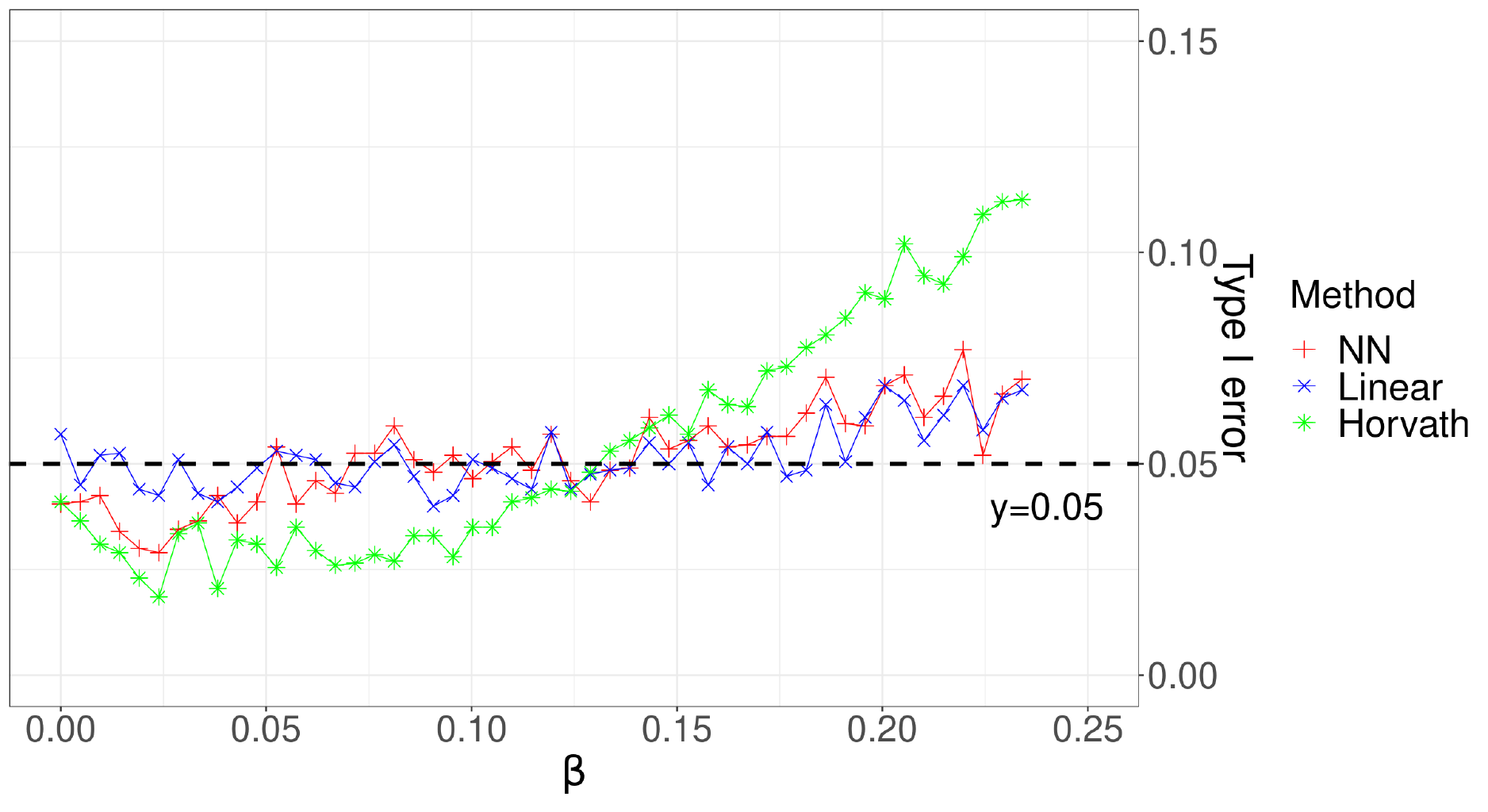}
        \caption{}
    \end{subfigure}
    \caption{\small Panel (a): Type I errors for neural network (red), linear classifier (blue), and \citet{hor20}'s method (green) assuming that the parameters $\beta \in \mathcal{B}_{\text{test}}$ are known where $\lvert \mathcal{B}_{\text{test}} \rvert = 50$ and the other two parameters of a Mat\'ern covariance function are fixed, $\sigma^2 = 1$ and $\nu = 0.5$;
        Panel (b): Same as (a) when the parameters $\beta$ on the $x$-axis and $\sigma^2$ are estimated by maximum likelihood estimates and the smoothness parameter is fixed to $\nu=0.5$. The black dashed horizontal line in each panel represents the 5\% Type I error.}\label{type1err:sce1:sce2}
\end{figure}

First, we compare the Type I errors for the method in \citet{hor20}, the linear and the neural networks classifiers assuming that the true $\beta$s in $\mathcal{B}_{\text{test}}$ are known, in order to calibrate the testing data points with a suitable cut-off value from the pre-computed kernel regressions. In practice, the true values of $\beta$ are unknown and require estimation, so in order to assess the Type I errors in a real case, we estimate $\beta$ and $\sigma^2$ simultaneously with fixed $\nu = 0.5$ using the software  \texttt{ExaGeoStatR} \citep{exageostatR}, which allows a unified, high-performance parallel system designed to optimize a covariance-based Gaussian likelihood for spatial data. With the help of advanced high performance dense linear algebra libraries, \texttt{ExaGeoStatR} offers exact solutions for calculating the inverse of the covariance matrix and its determinant, which are necessary for evaluating the Gaussian log-likelihood. The optimization step in \texttt{ExaGeoStatR} relies on the Bound Optimization BY Quadratic Approximation (BOBYQA) method, which is a numeric, global, derivative-free and bound-constrained optimization algorithm \citep{powell2009bobyqa}, such that we can obtain faster and more accurate estimation than brute force methods. Figure \ref{type1err:sce1:sce2} illustrates the resulting Type I errors for both the cases of known and unknown parameters. In the first case (known parameters), our adaptive cut-off methods have approximately nominal 5\% Type I error rates for all $\beta$ values (see the red and blue lines in Figure \ref{type1err:sce1:sce2}) while \citet{hor20}'s method has unstable Type I error rates as the dependence parameter varies (see the green lines in Figure \ref{type1err:sce1:sce2}). In the second scenario (unknown parameters), the outcomes are still comparable to those of known parameters although we utilize estimated $\beta$s instead of the true values.

In Section B of the supplement, we discuss the case where the parameter $\nu$ is misspecified. Specifically, we train the linear and neural network models using the data generated with $\nu=1$, while the actual test data are generated with $\nu=0.5$, and vice versa. The misspecification of $\nu$ significantly worsens the size of tests because the value of $\beta_{\text{max}}$, which controls the size of a test, is computed based on the wrong $\nu$, so incorrect cut-off functions are derived (see Figure \ref{cutoff:nu05} and Figure S1 in the supplement). In real-world scenarios, $\nu$ has to be estimated along with the linear models and neural networks. In Section \ref{sec:app}, we demonstrate how to practically calibrate the tests with an estimated $\nu$.

\subsubsection{Power comparison} 

\begin{figure}[b!]
    \centering
    \begin{subfigure}[t]{0.49\textwidth}
        \centering
        \includegraphics[height=5cm, width=7.2cm]{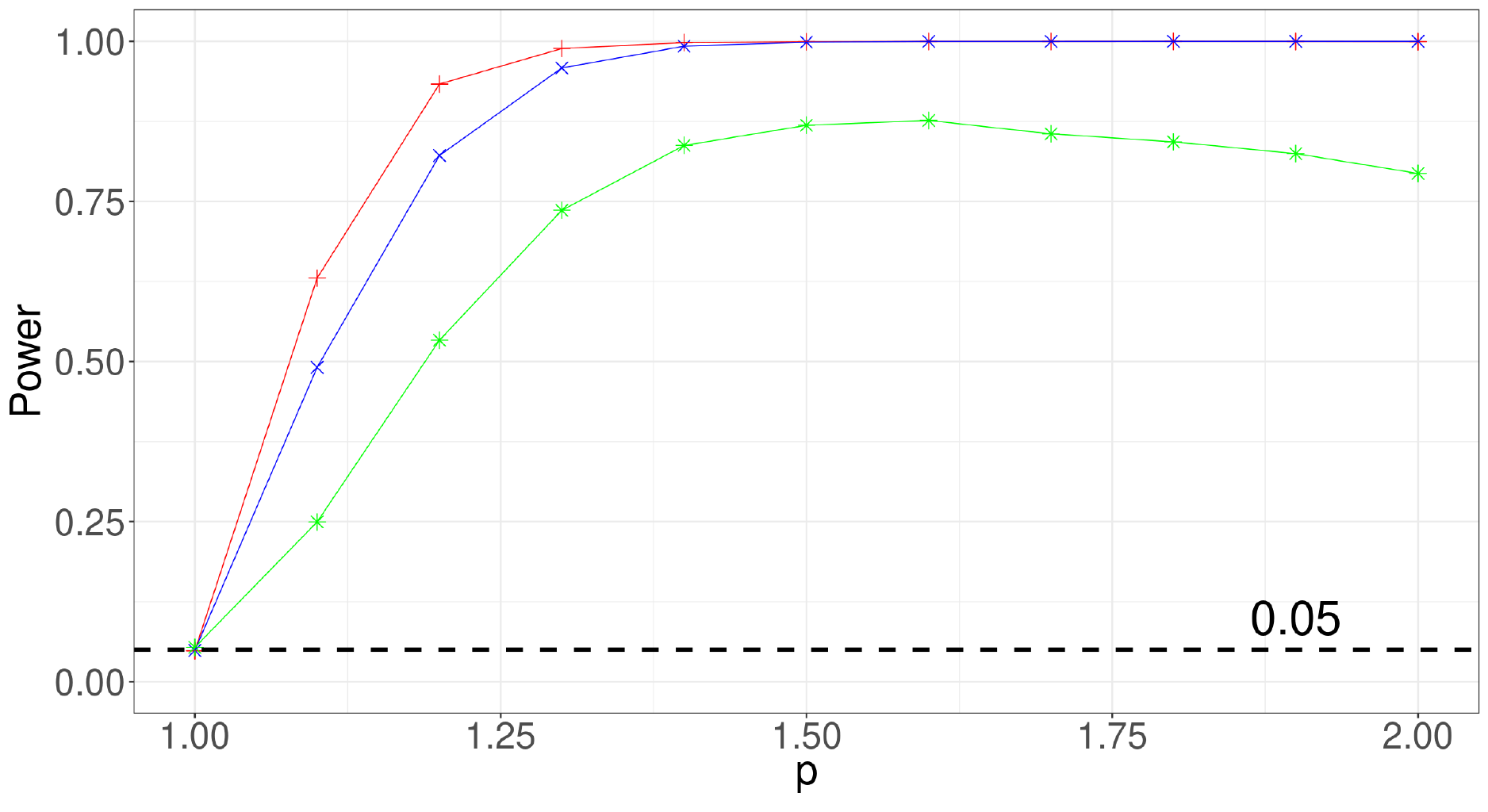}
        \caption{}
    \end{subfigure}%
    ~
    \begin{subfigure}[t]{0.49\textwidth}
        \centering
        \includegraphics[height=5cm, width=8.5cm]{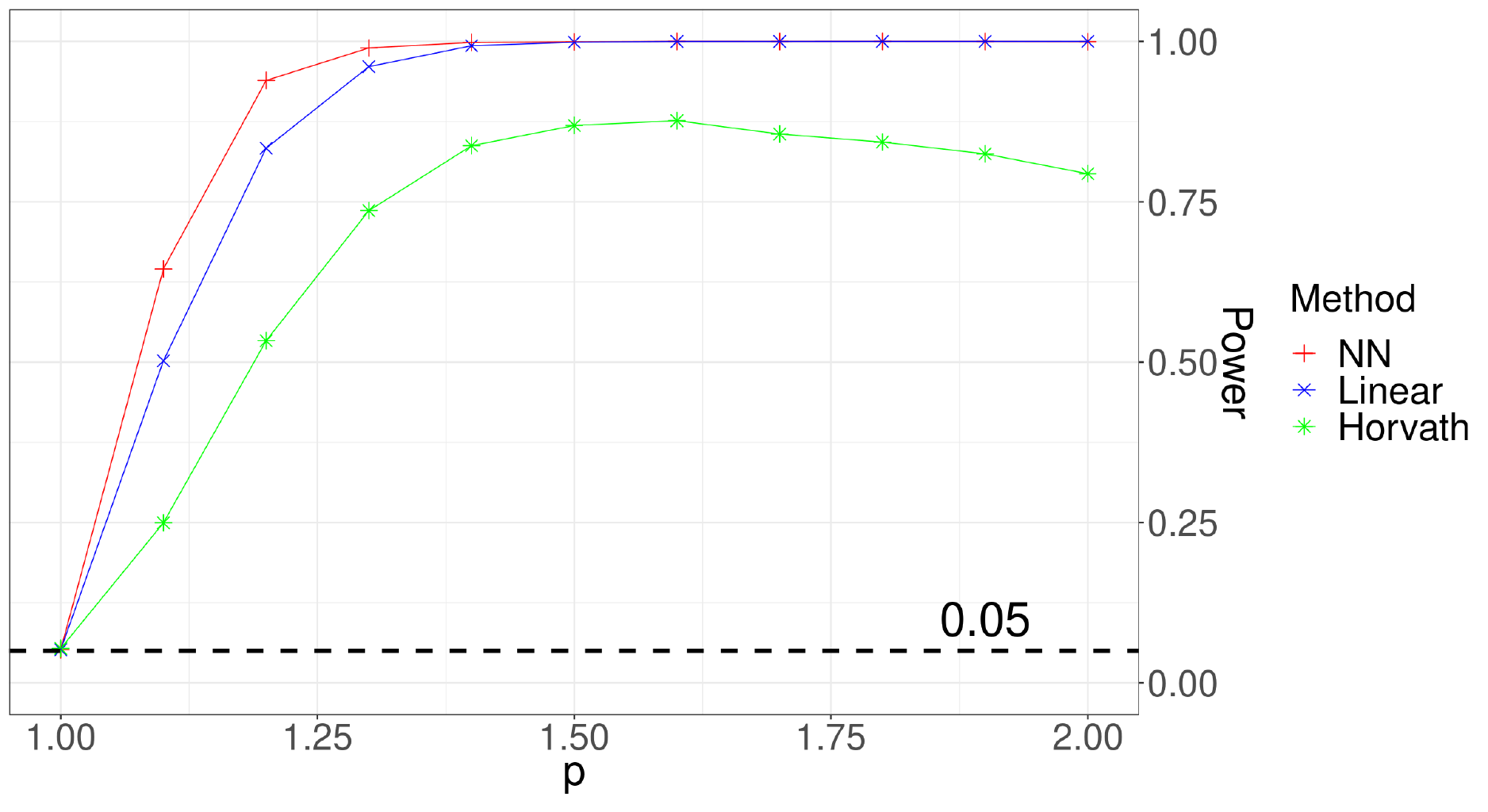}
        \caption{}
    \end{subfigure}
    \caption{\small Panel (a): Averaged powers across all choices of $\beta \in \mathcal{B}_{\text{test}}$ for neural network (red), linear classifier (blue), and \citet{hor20}'s method (green) given a value of exponent $p$ (i.e., non-normality parameter) on the $x$-axis assuming that the parameters $\beta \in \mathcal{B}_{\text{test}}$ are known where $\lvert \mathcal{B}_{\text{test}} \rvert = 50$ and the other two parameters of a Mat\'ern covariance function are fixed, $\sigma^2 = 1$ and $\nu = 0.5$;
        Panel (b): Same as (a) when the parameters $(\beta, \sigma^2)$ are estimated by maximum likelihood estimates and the smoothness parameter is fixed, $\nu=0.5$. The black dashed horizontal line in each panel represents the power of 5\%. }\label{overall:power}
\end{figure}

In order to identify the best test, we need to assess the power under the alternative hypothesis $H_1$ while maintaining a predetermined Type I error rate $\alpha$. We compare powers for our proposed neural network model and linear aggregation with adaptive cut-off, along with the approach in \citet{hor20}. Figure \ref{overall:power} shows the power curves as a function of the departure from normality, measured by the exponent $p$. Each curve is computed as an average across all choices of dependence parameters $\beta \in \mathcal{B}_{\text{test}}$ assuming that they are known (See Panel (a)) or estimated (See Panel (b)).

It is readily apparent that the neural network classifier achieves the highest power for all choices of $p \in \{1.1, 1.2, \dots, 2.0 \}$. Also, our adaptive cut-off method has higher power as the non-normal distribution's tails become heavier (with larger $p$). Here, neural networks perform only slightly better than linear combinations. The use of only six inputs can be one reason for the slight improvement in this case. It is expected that the accuracy of neural networks would be enhanced if a larger number of inputs are employed.

\subsubsection{Sensitivity analysis} \label{sim:sensitivity}

\begin{figure}[h!]
\vspace{1cm}
    \centering
    \begin{subfigure}[t]{0.49\textwidth}
        \centering
        \includegraphics[height=5cm, width=7.2cm]{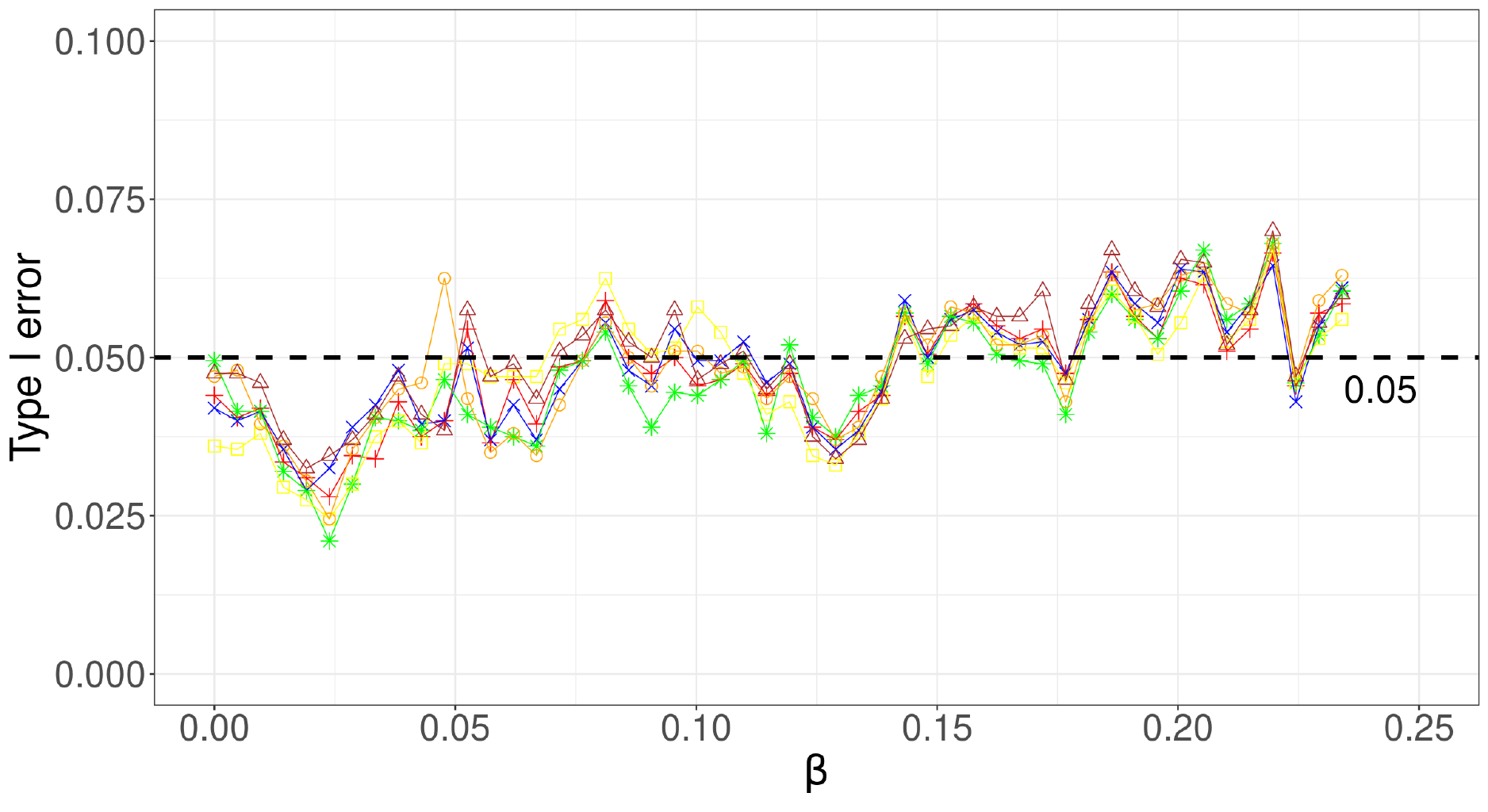}
        \caption{}
    \end{subfigure}%
    ~
    \begin{subfigure}[t]{0.49\textwidth}
        \centering
        \includegraphics[height=5cm, width=8.5cm]{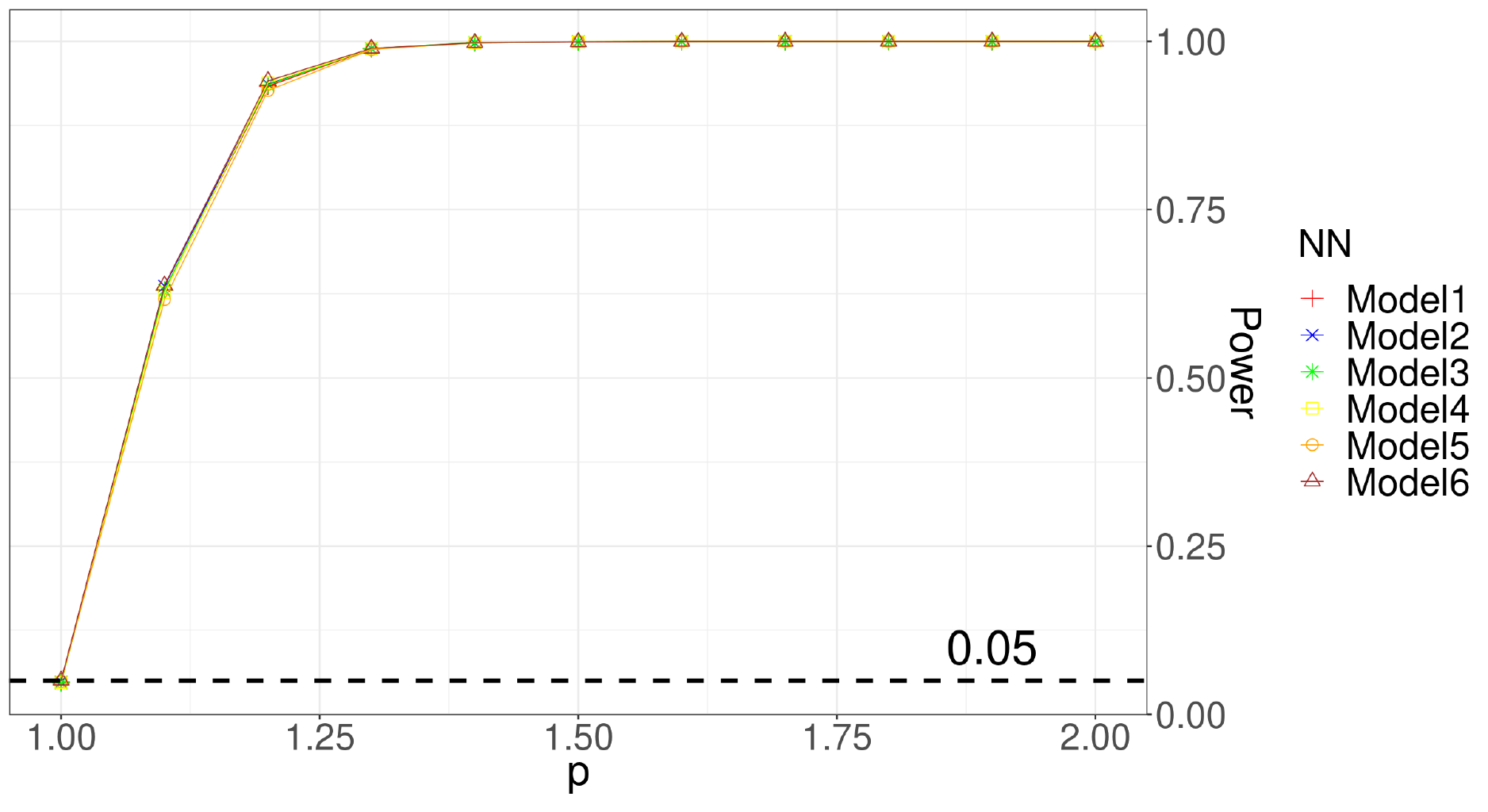}
        \caption{}
    \end{subfigure}
    \caption{\small (a): Type I errors for various architectures of neural networks---Model 1 (red), Model 2 (blue), Model 3 (green), Model 4 (yellow), Model 5 (orange), and Model 6 (brown)---assuming that the parameters $\beta \in \mathcal{B}_{\text{test}}$ are known where $\lvert \mathcal{B}_{\text{test}} \rvert = 50$; (b): Overall powers for various architectures of neural networks   computed as an average over all values of $\beta \in \mathcal{B}_{\text{test}}$. For both panels, the other two parameters of the Mat\'ern covariance function are fixed to $\sigma^2 = 1$ and $\nu = 0.5$ and the black dashed horizontal lines represent $y=0.05$.} \label{type1err:sensitivity}
\end{figure}

We perform a sensitivity analysis with respect to the choice of depth $L$, width $(n_1, n_2)$, and dropout rate of the neural network. First, we consider the same drop-out rate of 0.3 but different number of layers and nodes: 1) three hidden layers with $(n_1, n_2, n_3) = (256, 128, 64)$; 2) two hidden layers with $(n_1, n_2) = (32, 16)$; and 3) one hidden layer with $n_1 = 128$. Second, we use the same number of layers and nodes as in Section \ref{sim:result} but different drop-out rates, 0.6 or 0.1. Hence, we have a total of six distinct network structures, including the original one, and the results are summarized in Table \ref{table:various:networks}.
\begin{table}[h!]
\centering 
\caption{\small Summary of different network architectures: Model 1 is the original network we used in Section \ref{sim:result}. The drop-out rate in Models 2, 3, and 4 is identical to that of the original model, however, they differ in their network structures. Models 5 and 6 have modified drop-out rates with the same structure as the original one.} \label{table:various:networks}
\begin{tabular}{ccccccc}
  \hline
 & Model 1 & Model 2 & Model 3 & Model 4 & Model 5 & Model 6 \\\hline 
 \# of layers & 2 & 3 & 2 & 1 & 2 & 2 \\
 \# of nodes & (256, 128) & (256, 128, 64) & (32, 16) & (128) & (256, 128) & (256, 128) 
 \\
 Drop-out & 0.3 & 0.3 & 0.3 & 0.3 & 0.6 & 0.1 \\
\hline
\end{tabular}
\end{table}
We also recompute the Type I error and power in Figure \ref{type1err:sce1:sce2}-(a) and Figure \ref{overall:power}-(a) for all models. The results, shown in Figure \ref{type1err:sensitivity}, show how all six networks display a very similar pattern.

\section{Testing Normality for Global Climate Data}\label{sec:app}

\subsection{Motivation}
Climate change is bound to affect both natural and human systems, with varying outcomes depending on the region, economic sector, and time. The magnitude and range of future climate does not only rely on the dynamics of the Earth's system but also on scenarios of socio-economic developments \citep{ipcc22}. Computer models or simulators are the standard tool to understand and quantify future changes in the climate, as well as their social, political and economic effects. The high complexity, spatial and temporal resolution of modern climate models make it impossible to explore future climate for a fully exhaustive range of scenarios, as every simulation puts a considerable strain on the computational and storage resources of an institution's cyberinfrastructures \citep{hua23}. As such, sensitivity analysis is limited to a selected set representative of physical parametrizations and scenarios, and uncertainty quantification can be performed partially at best. Statistical surrogates, or emulators \citep{sac89,ken01} are then routinely trained on a small set of available simulations, and then used to provide a considerably faster (yet approximate) assessment of the behavior of (some variables at some spatio-temporal resolutions of) a climate model \citep{cas13,castruccio2014statistical,cas14}. A useful simplifying assumption for climate emulation is that of Gaussianity, which at some level of spatial and/or temporal aggregation is more or less explicitly assumed to be valid owing to the central limit theorem. The presence of spatial and temporal dependence within the data, however, makes it challenging to formally assess this assumption. Testing for normality in this framework is therefore of high relevance as it would provide indications as to which modeling strategy would be more appropriate: a Gaussian process emulator \citep{sac89} or more complex trans-Gaussian \citep{jeo19,tag20} or latent Gaussian models \citep{zha23}. In this application, we make use of our adaptive cut-off method to assess normality of a widely used collection of climate simulations under different levels of aggregation.

\subsection{CMIP6 data}

We focus on the data from the Coupled Model Intercomparison Project Phase 6 (CMIP6, \cite{eyr16}), the reference collection of simulations (\textit{ensemble}) of the Intergovenmental Panel on Climate Change Assessment Report 6 \citep{juckes2020cmip6} and in particular on the MIROC-ES2L model \citep{hajima2020development} given its complete record of simulations. We consider on monthly near surface air temperature data (at 2 meters above the ground level, in Celsius) under SSP245, an intermediate scenario in terms of global mean temperature increase and degree of global socio-economic collaboration throughout the 21st century  \citep{van2014new}. The data set comprises $T=12\times 86=1032$ time points (all months in 2015--2100) on a regular $2.79^\circ \times 2.81^\circ$ latitude and longitude grid, for a total of $M=64\times 128=8192$ locations. We denote the temperature as $Y_t(\mb{s}_i)$ at location $i = 1, \dots, M$ and time point $t = 1, \dots, T$. Before assessing normality, we provide a model for the trend and the temporal dependence, which need to be removed before applying our proposed methdology. 

\subsection{Modeling trend and temporal dependence}

We consider the following additive spatio-temporal autoregressive moving average (ARMA)-like model:\\
\begin{subequations}\label{eqn:modelcmip}
\begin{eqnarray}
\label{mean-model}
    Y_t(\mb{s}_i) & = & \mu_{r(t)}(\mb{s}_i) + \epsilon_t(\mb{s}_i),\\    \epsilon_t(\mb{s}_i) & = & \sum_{j=1}^p \psi_{j;i} \epsilon_{t-j}(\mb{s}_i) + \sum_{k=0}^q \theta_{k;i} \eta_{t-k}(\mb{s}_i).\label{eq:arma}
\end{eqnarray}
\end{subequations}
where $\theta_{0;i}=1$, $\mu_{r(t)}(\mb{s}_i)$ is the monthly trend with indices $r(t) \in \{0, \dots, 11\}$ representing the remainder when $t$ is divided by $12$ and $\eta_t(\mb{s}_i)$ is a zero-mean residual uncorrelated in time. Further, we assume that $\text{Var}\{ \epsilon_t(\mb{s}_i) \} = \sigma_{r(t)}^2(\mb{s}_i)$ for $t = 1, \dots, T$, i.e., there is a month-specific variance. For each location independently, both mean and variance are estimated in a non-parametric fashion with a moving window estimator:
\begin{align*}
    \widehat{\mu}_{r(t)}(\mb{s}_i) &= \frac{1}{\lvert A_{r(t)} \rvert} \sum_{t \in A_{r(t)}}Y_t(\mb{s}_i), \quad
    \widehat{\sigma}_{r(t)}^2(\mb{s}_i) = \frac{1}{\lvert A_{r(t)} \rvert} \sum_{t \in A_{r(t)}} \left\{ Y_t(\mb{s}_i) - \widehat{\mu}_{r(t)}(\mb{s}_i) \right\}^2,
\end{align*}
where $A_{r(t)} =\{t : \text{$t$ mod 12} = r(t) \}$. The average $R^2$ across all locations is $0.80$ with standard deviation $0.21$ and 89\% values of $R^2$ are greater than 0.5, which is better than harmonic regression (performed in the supplementary material). We then remove the trend and variance by computing the standardized residuals as:
\[
    \widehat{\epsilon}_t(\mb{s}_i) = \frac{Y_t(\mb{s}_i) - \widehat{\mu}_{r(t)}(\mb{s}_i)}{\widehat{\sigma}_{r(t)}^2(\mb{s}_i)}.
\]

Finally, for each location, we perform inference on the ARMA model \eqref{eq:arma} on $\widehat{\epsilon}_t(\mb{s}_i)$ using the R package \texttt{forecast} \citep{hyndman2008automatic}, with the orders $p$ and $q$ selected via Bayesian information criterion (BIC). Once the model orders are identified, the model parameters $\psi_{j;i}$ and $\theta_{k;i}$ are estimated by maximum likelihood inference and we use them to compute the residuals $\widehat{\eta}_t(\mb{s}_i)$ as estimates of our target quantity $\eta_t(\mb{s}_i)$. 

Intuitively, the normality assumption for the air temperature data would be violated due to the occurrence of exceptional temperatures at certain locations, resulting in heavier tail probabilities compared to a Gaussian distribution. Hence, it might not be preferable to employ the normality assumption for modeling the original temperature data. In this regard, we are interested in assessing the impact of spatial aggregation on the normality of $\widehat{\eta}_t(\mb{s}_i)$. To simplify the notation, we will abuse the notation and use the same expression for the residuals at different levels of spatial aggregation.

\subsection{Data aggregation}\label{data:agg}

\begin{figure}[h]
    \centering
    \begin{subfigure}[t]{0.48\textwidth}
    \caption{}
        \centering
        \includegraphics[height=4.8cm, width=7.2cm]{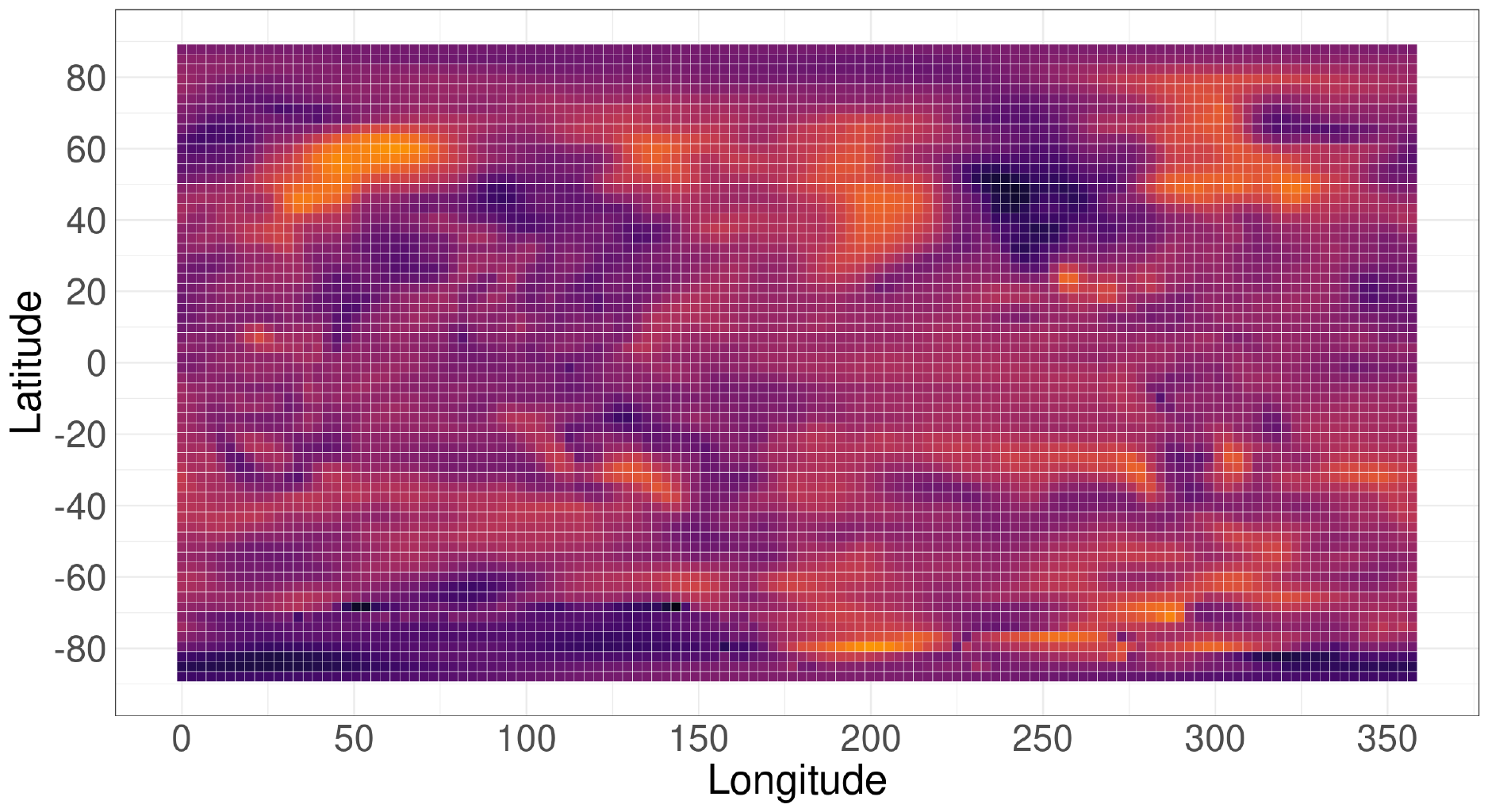}
    \end{subfigure}%
    ~\hspace{.2cm}
    \begin{subfigure}[t]{0.48\textwidth}
    \caption{}
        \centering
        \includegraphics[height=4.8cm, width=8.5cm]{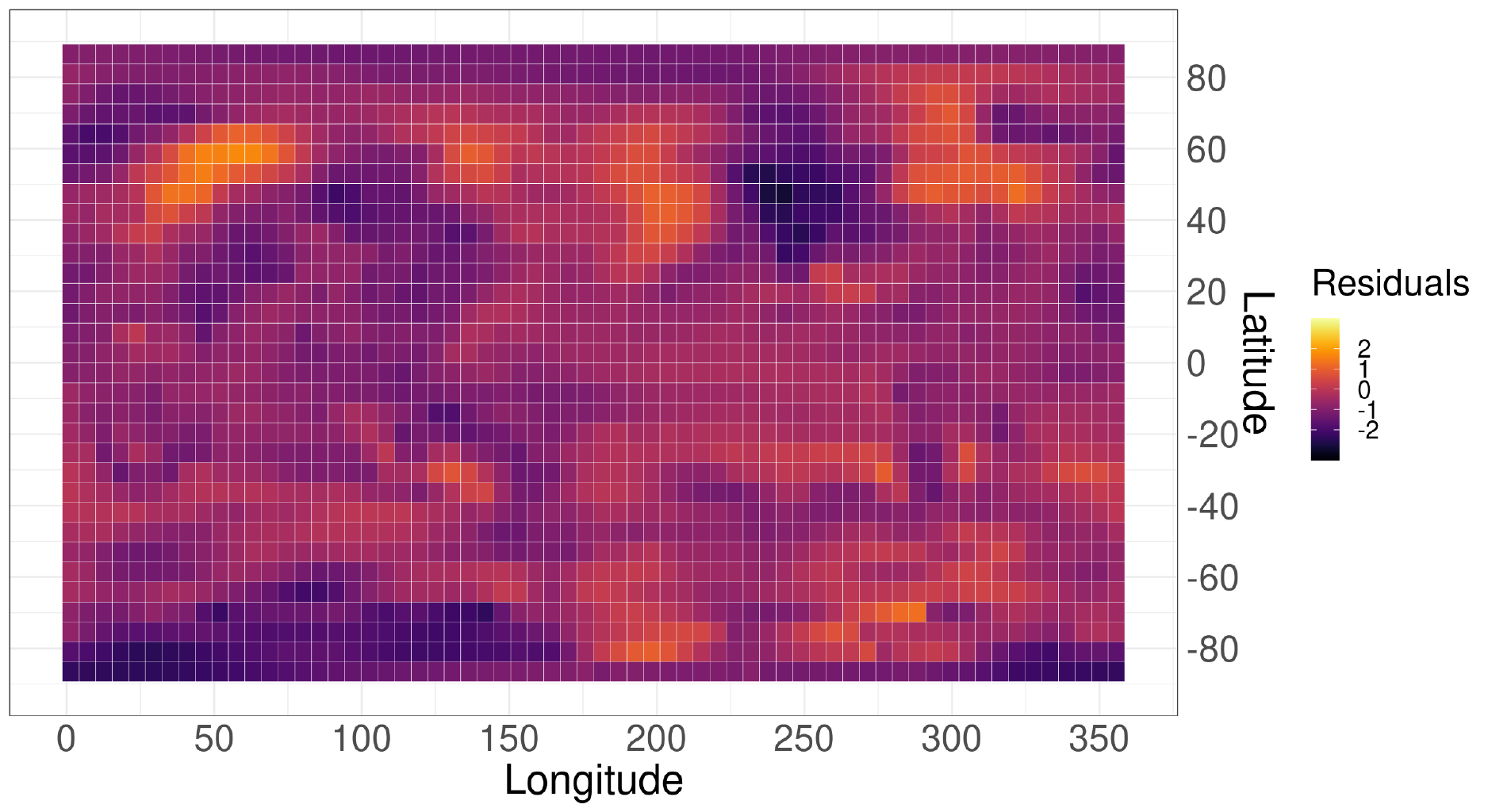}
    \end{subfigure}
    ~
    \begin{subfigure}[t]{0.48\textwidth}
    \caption{}
        \centering
        \includegraphics[height=4.8cm, width=7.2cm]{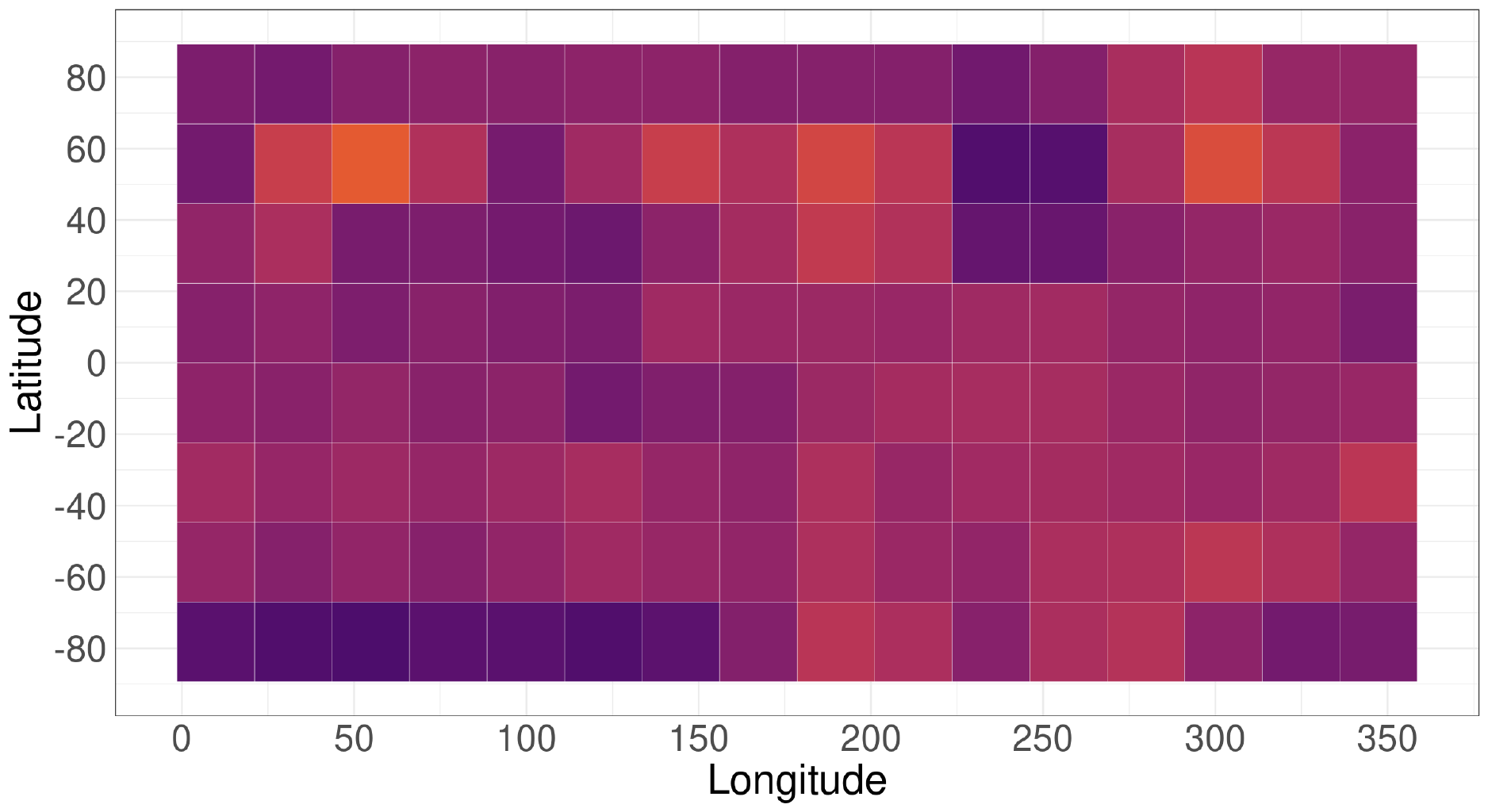}
    \end{subfigure}
    ~\hspace{.2cm}
    \begin{subfigure}[t]{0.48\textwidth}
    \caption{}
        \centering
        \includegraphics[height=4.8cm, width=8.5cm]{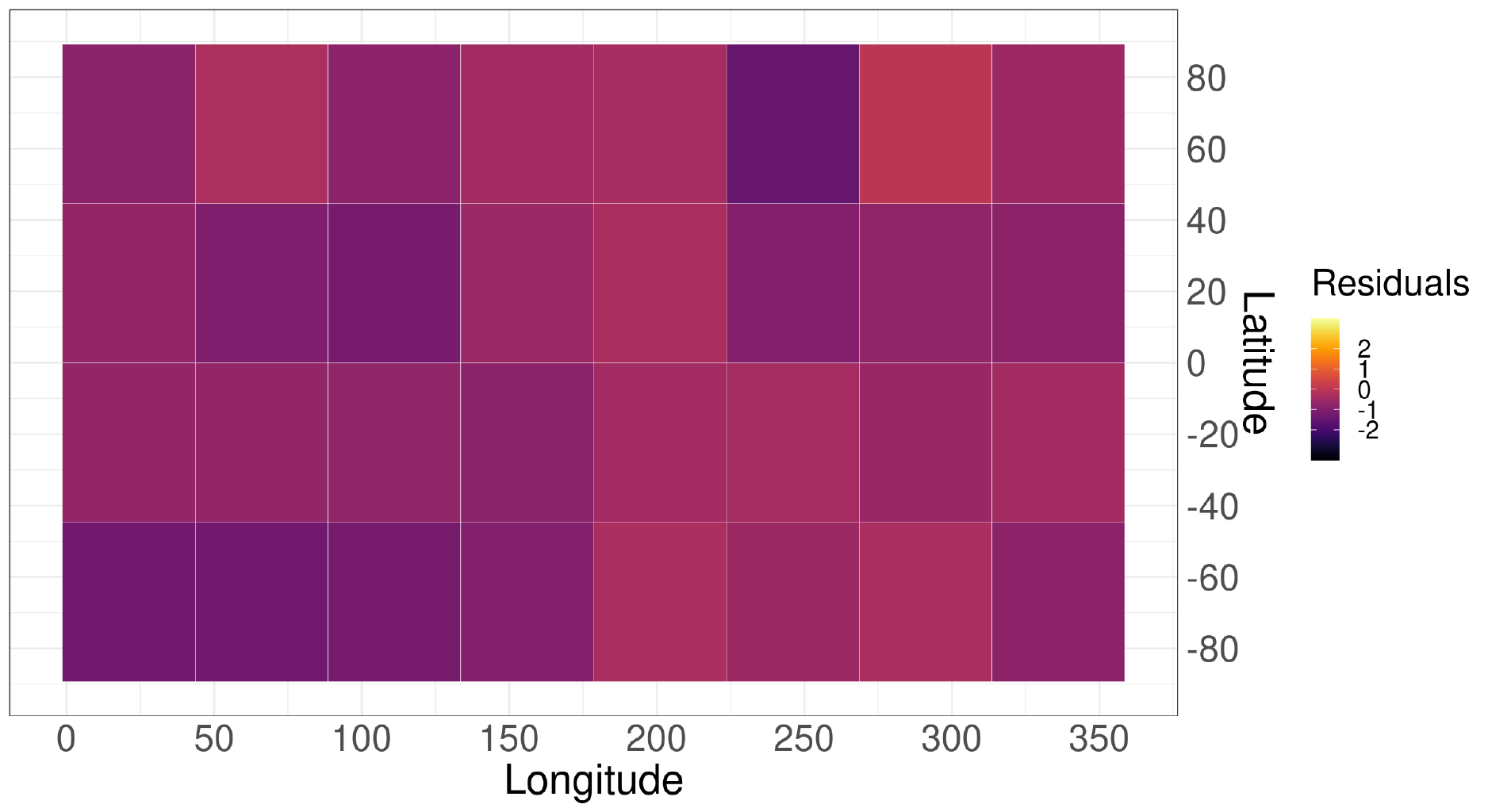}
    \end{subfigure}
    
    \caption{\small Standardized emulator residuals $\hat{\eta}_t(\mb{s}_i)$ in January 2015 for different levels of spatial aggregation. (a): original grid resolution; (b): 4 observations in each square of size $2 \times 2$; (c): 64 observations in each square of size $8 \times 8$; (d): 256 observations in each square of size $16 \times 16$. } \label{heatmap}
\end{figure}

The emulator residuals $\hat{\eta}_t(\mb{s}_i)$ are likely not normal at the native grid resolution, as it is expected that some locations will have unusual temperatures with heavier-than-normal tails. However, some degree of spatial aggregation should result in more normal residuals, and we aim at formally testing this assumption with our proposed approach. We partition the pixels (locations) into smaller squares and compute the mean of the estimated residuals, $\hat{\eta}_t(\mb{s}_i)$, within each square. We choose the square of sizes $2 \times 2$, $4 \times 4$, $8 \times 8$, and $16 \times 16$ such that the corresponding aggregated data have the number of locations $M=2048, 512, 128, 32$, respectively. Figure \ref{heatmap} shows the map of the estimated residuals in January 2015 at all four different levels of aggregation. 

\subsection{Calibration of classifiers}
\label{nn:calibration}

 First of all, we simulate the data from a Gaussian distribution using the Mat\'ern covariance in \eqref{eq:matern} with $\nu \in \mathcal{N}_{\text{train}} = \{0.5, 1.0, 1.5, 2.0, 2.5, 3.0 \}$  covering rough to smooth spatial processes, $\sigma^2 = 1$, and $\beta \in \mathcal{B}_{\text{train}} = \{0, \dots, \beta_{\max} \}$, thereby covering independence to strong dependence and $n_{\beta;\text{train}} = \lvert \mathcal{B}_{\text{train}} \rvert = 30$ as in Section \ref{sec:sim}. The range parameter bound $\beta_{\max}$ depends on the choice of $\nu$ and the spatial domain. Since in the case of a unit square we had the effective range of 0.7 corresponding to the strong dependence, for the domain here, we rescale it using the following ratio: effective range/maximum distance = $0.7/\sqrt{2}$,
where the maximum distance and the effective range are 6307 km  and 12742 km, respectively, in  chordal distance for all levels of aggregation. The different values of $\beta_{\text{max}}$ across different choices of the smoothness parameter $\nu$ are shown in Table S2 of the supplementary materials. Here, we emphasize that we train six pairs of neural networks and linear classifiers for each value of $\nu$ and every testing data point will be assigned to one of the six based on the estimated value of $\nu$.

\begin{figure}[t!]
    \centering
    \begin{subfigure}[t]{0.49\textwidth}
        \centering
        \includegraphics[height=5cm, width=7.0cm]{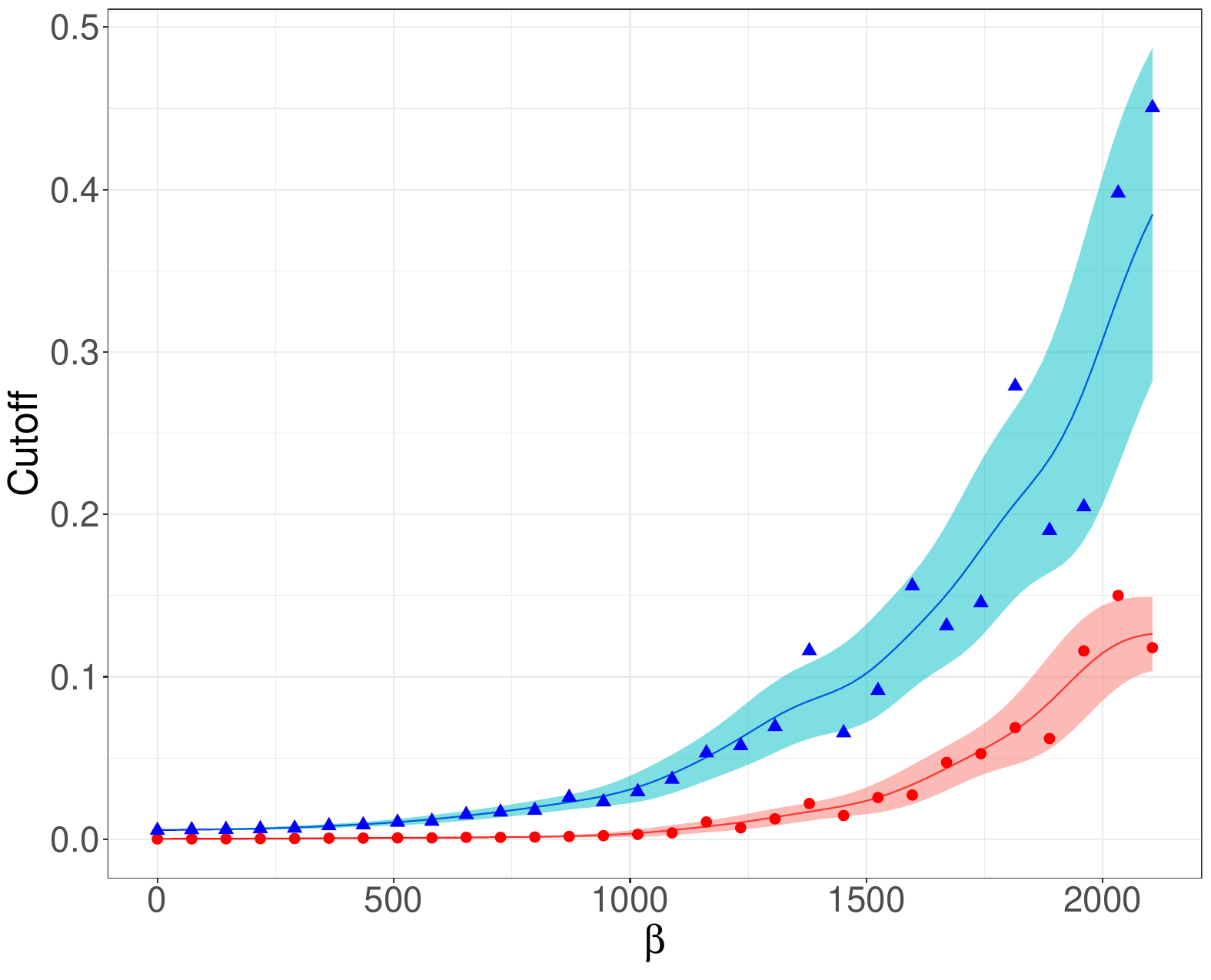}
        \caption{}
    \end{subfigure}%
    ~
    \begin{subfigure}[t]{0.49\textwidth}
        \centering
        \includegraphics[height=5cm, width=8.5cm]{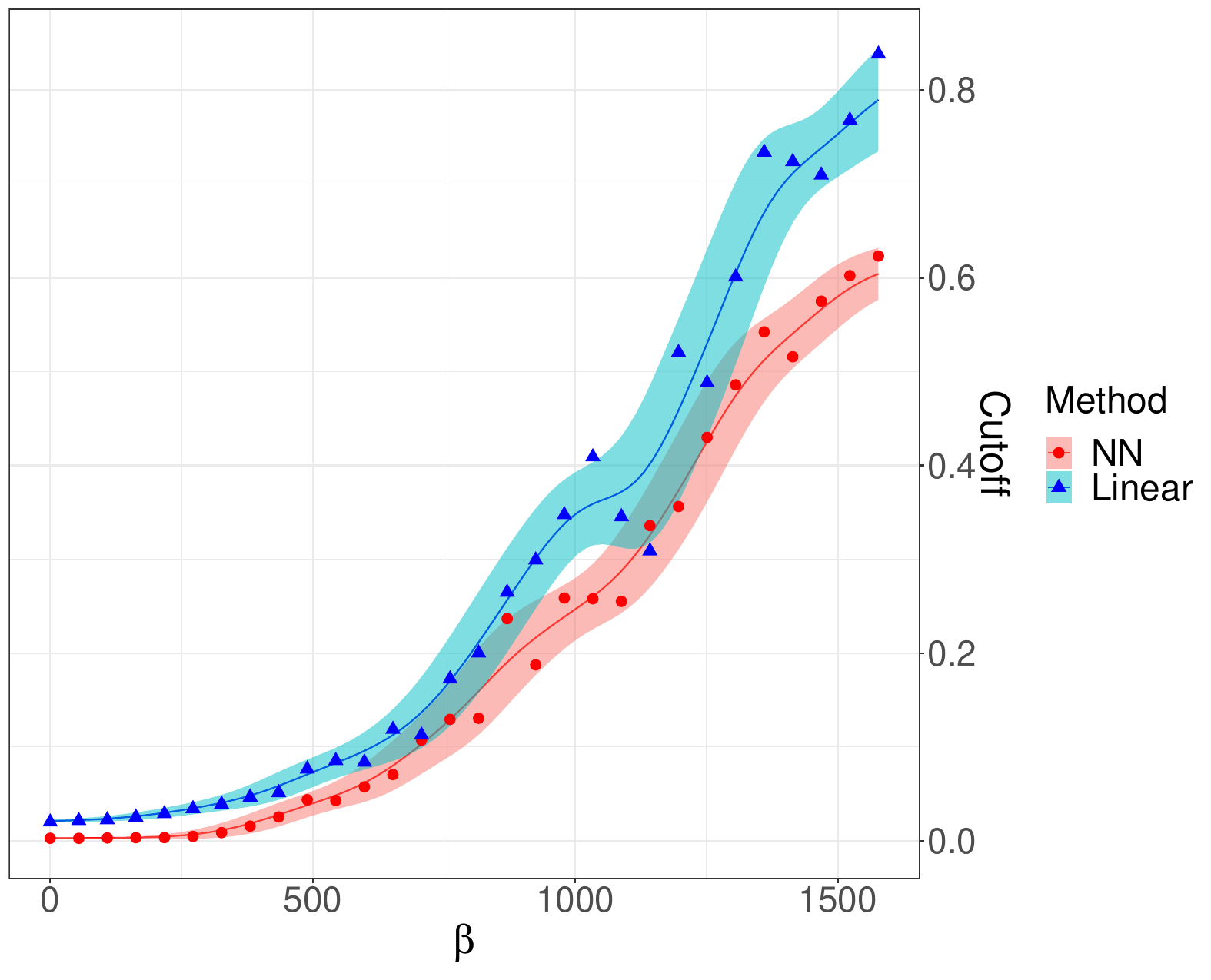}
        \caption{}
    \end{subfigure}
    \caption{\small Application (native grid resolution): Non-parametric Gaussian kernel regressions as defined in \eqref{eq:kernel} with a bandwidth $h=0.3$ for neural network (red) and linear (blue) classifiers. On the $x$-axis are the range parameter $\beta$ of the Mat\'ern covariance \eqref{eq:matern}, while on the $y$-axis the predicted cut-off and corresponding pointwise 95\% confidence interval are represented by solid lines and bands, respectively. Since the residuals are normalized, we set $\sigma^2 = 1$, while we have the smoothness parameter equal to (a) $\nu=0.5$ and (b) in $\nu=1.0$.}
\label{cutoff:realdata}
\end{figure}

For non-normal data, the same transformation as in Section \ref{sec:sim} is used with $p \in \mathcal{P}_{\text{train}} = \{1.2, 1.4, 1.6, 1.8\}$. We draw $n_{\text{sample}} = 200$ sample points for each setup such that we have $n_{\nu;\text{train}} \times n_{\beta;\text{train}} \times n_{p;\text{train}} \times n_{\text{sample}} = 144,000$ non-normal data points and the same amount of normal data points where $n_{\nu;\text{train}} = \lvert \mathcal{N}_{\text{train}} \rvert 
 = 6$. Calibration is performed with the simulated normal and non-normal data and the resulting cutoff functions for each value of $\nu \in \mathcal{N}_{\text{train}}$ are obtained using non-parametric kernel regression as illustrated in Figure \ref{cutoff:realdata}. 

For the structure of neural networks, the number of hidden layers is $L=2$ with $n_1=256$ and $n_2=128$ nodes and we use $m=5$ inputs among those we used in Section \ref{sec:sim}. We do not use the Shapiro--Wilk test because the number of locations at the original resolution, $M = 8192$, exceeded the maximum allowed by the \texttt{R} implementation of the test (see the discussion on the methods about reliability of the test for large $M$ in Section \ref{sec:indiv:tests}). To determine suitable neural network and linear classifiers and corresponding cut-off values for each time point $t=1, \dots, T$, we estimate the Mat\'ern parameters $(\sigma^2, \beta, \nu)$ simultaneously given the location information with chordal distances and the spatial residuals $\left(\hat{\eta}_t(\mb{s}_1), \dots, \hat{\eta}_t(\mb{s}_M) \right)^\top$ using the package \texttt{ExaGeoStat} \citep{abdulah2018exageostat} which relies on BOBYQA optimization \citep{powell2009bobyqa}. Then, each testing data vector is allocated to a trained neural network and a linear classifier according to the closest approximation of the estimated smoothness parameter. For example, if the estimated smoothness parameter for a data vector is $\hat{\nu} = 0.8$, we use the neural network and linear classifier calibrated with $\nu=1$, if $\hat{\nu} = 0.3$, we use the neural network and linear classifier calibrated with $\nu=0.5$. 

\subsection{Test results}

For the different levels of data aggregation we perform the calibration as detailed in Section~\ref{nn:calibration} and compute the rejection rates across all time points ($T=1032$). The results are shown in Table \ref{rejection:ratio}. As expected by the central limit theorem, as the spatial aggregation increases, both the neural network and the linear test highlight that the residuals become more normally distributed. Indeed, at native resolution the normality tests are rejected for more than 95\% of time points for both classifiers, while higher levels of aggregation decrease the rejection rates down to approximately 20\%. The neural network model is overall less favorable towards the normality assumption, and the discrepancy between the two approaches is  slightly higher when the degree of spatial aggregation is moderate ($M=512)$. As we expected, the rejection rate is very high with the original resolution of the temperature data, and interestingly, the rejection rate is still high with the moderate level of aggregation, therefore flagging the normality assumption as generally inappropriate. 
This can likely be attributed to a large number of time points ($T=1032$), which result in high power of a normality test against any alternative distribution.

\begin{table}[t!]
\centering
\caption{\small Rejection rates for the estimated residuals of the emulator \eqref{mean-model} for the neural network and linear normality testing approach. The results are shown across the different level of spatial aggregation.}
\label{rejection:ratio}
\begin{tabular}{cccccc}
  \hline
 Rejection rate & All locations ($M=8192$)  & $M=2048$ & $M=512$ & $M=128$ & $M=32$ \\
  \hline 
  NN & 0.994 & 0.967 & 0.845 & 0.532 & 0.227 \\
  Linear & 0.958 & 0.924 & 0.735 & 0.511 & 0.191 \\
   \hline
\end{tabular}
\end{table}


\section{Discussion and Conclusion}\label{sec:concl}

We proposed a new test for dependent data to test Gaussianity by merging the test statistic of individual normality tests (which may or may not assume dependence) via neural networks. By means of a simulation study, we have shown how the proposed approach results in higher power than individual tests as well as a linear aggregation of the tests. Our application for temperature data highlighted how increasing the level of spatial aggregation results in more normal data, as could be expected from the central limit theorem. 

The proposed approach has been applied to normality test for dependence data, but its extent is far more general. In fact, other marginal distributions can be tested: a generalized extreme value distribution can be assessed for maxima at different levels of temporal aggregation, or skew-normality for high resolution weather data. Such approach could also be generalized to multivariate data to test marginal univariate properties. 

While the proposed approach represents a significant step forward in assessing Gaussianity under dependence, it comes with several caveats that a practitioner must be aware of. Firstly, the method must assume a given structure of spatial dependence, so the reliability of the results are inextricably linked with the assumptions associated with it, most noticeably isotropy and stationarity. While these assumptions may be hard to defend for the original data, the focus on residuals would at least partially justify the spatial structure. Additionally, the proposed method depends on a prespecified type of alternative hypothesis, in this case a non-Gaussian power transformation, and this may or may not be a good alternative hypothesis depending on the application.


\section*{Supporting Information and Data Availability}
The code for this work is available at  
$
\texttt{https://github.com/stat-kim/adaptive-cutoff}
$
The data that support the findings of this study are openly available as part of the Large Ensemble project at the National Center for Atmospheric Research at www.earthsystemgrid.org.

\section*{Acknowledgments}

We would like to thank Brian Greco for insightful discussions. This research was supported by the King Abdullah University of Science and Technology (KAUST). 

\baselineskip 16pt

\bibliographystyle{jasa3}
\bibliography{biblio}
\end{document}